\documentclass[reprint
]{revtex4-2}

\usepackage{graphicx}
\usepackage{subcaption}
\usepackage{dcolumn}
\usepackage{bm}
\usepackage{hyperref}
\usepackage{xcolor}
\usepackage{amsmath}

\begin{document}

\title{Microwave Studies of Single Crystal TeO2 at Cryogenic Temperatures}

\author{Timothy Holt}
\affiliation{ARC Centre of Excellence for Dark Matter Particle Physics, Department of Physics, University of Western Australia, 35 Stirling Highway, Crawley WA 6009, Australia}

\author{Maxim Goryachev}
\email{maxim.goryachev@uwa.edu.au}
\affiliation{ARC Centre of Excellence for Dark Matter Particle Physics, Department of Physics, University of Western Australia, 35 Stirling Highway, Crawley WA 6009, Australia}

\author{William Campbell}
\affiliation{ARC Centre of Excellence for Dark Matter Particle Physics, Department of Physics, University of Western Australia, 35 Stirling Highway, Crawley WA 6009, Australia}

\author{Michael E. Tobar}
\affiliation{ARC Centre of Excellence for Dark Matter Particle Physics, Department of Physics, University of Western Australia, 35 Stirling Highway, Crawley WA 6009, Australia}

\date{\today}

\begin{abstract}
{We use whispering-gallery-mode analysis to characterise the microwave dielectric properties of single-crystal TeO$_2$ at cryogenic temperatures and compare its loss performance with other low-loss dielectric materials. Finite-element modelling is combined with measurements at room temperature, 4 K, and 20 mK to develop accurate cryogenic simulations and extract the anisotropic dielectric permittivities, giving $\varepsilon_\parallel=25.75\pm0.08$ and $\varepsilon_\perp=20.90\pm0.07$. Loss measurements reveal quality factors as high as $9\times10^6$ and minimum loss tangents approaching $3\times10^{-8}$, placing TeO$_2$ among promising low-loss dielectrics for cryogenic microwave applications. Electron-spin-resonance spectroscopy further indicates a clean spin environment, while identifying distinct spin systems consistent with the known properties of the crystal.}

\end{abstract}

\maketitle


\section*{\label{sec:level1}Introduction}

Precision experiments for quantum metrology and fundamental-physics tests often measure or search for extremely weak signals. These experiments require exceptional sensitivity and a detailed understanding of loss mechanisms in crystal substrates and resonators. Extensive work has characterised crystals such as sapphire for use in high-$Q$ resonators \cite{article2,Creedon2011}, with further studies extending to materials with related properties, including calcium tungstate (CaWO$_4$) and isotopically enriched silicon-28 ($^{28}$Si) \cite{ldsb-4dnr,10.1063/5.0224102,dbc7810e9767435d8ab8faba2b66db43}. Sapphire’s low cryogenic loss, high stability, and linear power dependence have enabled advances in precision clocks and related experiments. These experiments include high-precision tests of Lorentz invariance \cite{Nagel2025,Tobar2010}, local position invariance \cite{Tobar2013}, and quantum gravity \cite{Bushev2019}. Factors such as ease of fabrication, potential use in qubits, and distinctive optical and mechanical properties also make CaWO$_4$ and $^{28}$Si attractive materials for precision measurements and tests of fundamental physics. In these experiments, precise knowledge of impurities is crucial, as it allows weak signals to be cross-checked against known spectral features and helps assess their physical significance.

Whilst sapphire and its counterparts have widespread uses, it is important to study other materials for unique desirable properties to use in further experiments. Crystalline $\alpha$-TeO$_2$ (hereafter TeO$_2$) has a range of advantageous properties and is beginning to see use in several cryogenic experiments as the bulk material of bolometers \cite{ARNABOLDI2004775,Alessandria_2013}. The material's low dielectric loss at cryogenic temperatures makes it advantageous for supporting high-quality electromagnetic resonances, and its high dielectric permittivity values may enable the exploration of frequency ranges that crystals of comparable loss are excluded from \cite{MOUFOK2019102315,10.1063/1.3406135,PhysRevB.4.3736,4319236,10.1063/1.3595942}. Despite its increasing use and attractive properties, the cryogenic microwave characteristics of TeO$_2$ are relatively untested in comparison with materials like sapphire, with relevant microwave cryogenic studies remaining above temperatures of 10 K \cite{10.1063/1.1805717, 10.1063/1.1659223}. Additionally, TeO$_2$ possesses several novel characteristics that may enable its use in emerging technologies, and an in-depth knowledge of the crystal's structure and impurities is required for proper application in the relevant fields. 

One unique property of the crystal is its large double-beta-decay and high natural abundance of $^{130}$Te. This motivates TeO$_2$ neutrinoless double beta decay detection experiments such as CUORE and CROSS, aiming to test the Majorana nature of neutrinos \cite{SCHAFFNER201530, 10.1063/1.5031485, LCardani_2012, Avignone_2024,article,ARNABOLDI20102999,CASALI201744}. These experiments are aided by previously explored $\alpha$ energy discrimination techniques, allowing legitimate signals found with a TeO$_2$ detector to be verified. TeO$_2$ bolometers have also been considered for dark sector particle searches (such as axions and weakly interacting massive particles), and experiments like CUORE are secondarily outfitted for this purpose \cite{ARNABOLDI2004775,Alessandria_2013}. 

In addition to its relevance for fundamental physics, TeO$_2$ is also attractive for applied technologies. The crystal possesses several useful optoelectronic and optomechanical properties, including a large band gap, strong optical nonlinearity, and low optical and mechanical loss. These characteristics have led to its use in precision measurement systems and related technologies, including semiconductors, acousto-optic modulators, frequency converters, bolometers, and microwave-to-optical conversion schemes based on acoustic coupling \cite{10.1121/1.405652,doi:10.1126/science.abj4396,Shao:19,MOUFOK2019102315,10.1063/1.1657275,SCHAFFNER201530}. Of particular interest is its ability to bridge optoelectronic and optomechanical regimes, suggesting that TeO$_2$ could provide an alternative to materials such as lithium niobate in hybrid quantum systems \cite{Chiappina:23}. A detailed understanding of spin transitions within the crystal structure may also enable the use of TeO$_2$ in quantum technologies requiring long spin coherence times, including qubit substrates. This further motivates cryogenic spectroscopic studies of TeO$_2$ to expand its role in precision measurement and quantum-technology applications \cite{PhysRevLett.115.013601,MAZZOCCHI20191,article7}.

In this work, a Whispering-Gallery Mode (WGM) resonator is used to probe a TeO$_2$ sample at cryogenic temperatures, aiming to characterise a range of material properties. In Section \ref{sec:level2}, the WGM setup is introduced, explaining the mode analysis and spectroscopy methods utilised to produce the results. Section \ref{sec:level3} describes the mode characterisation methods, comparing room temperature measurements with Finite Element Modelling (FEM) simulations to accurately understand the cryogenic system. In Section \ref{sec:level4}, loss characteristics of TeO$_2$ are discussed using Quality factor Q and Loss Tangent tan$\delta$ results. Lastly, Section \ref{sec:level5} discusses Electron Spin Resonance (ESR) spectroscopy measurements and identifies spin systems within the crystal structure, then the final section concludes and looks towards the future of TeO$_2$ experiments.

\section{\label{sec:level2}Experimental Methods}


This work involves simulations and experiments for a cylindrical TeO$_2$ single crystal sample from Kinheng Crystal, with dimensions $D=32.05\pm0.05$ mm, $L=28.55\pm0.05$ mm and all sides polished. WGM solutions were analysed from 2-10 GHz to find mode Q-factors and eigenfrequencies, as well as the corresponding directionally dependent tan$\delta$ and electrical filling factor $p_i$ values. The comparison of experiment and simulation allows for several valuable insights into the crystal's properties, and additionally produces a cryogenic measurement of the crystal's dielectric permittivity tensor. ESR is also performed on identified high-Q WGMs, investigating Zeeman splitting of spin systems within the crystal lattice to aid in structural analysis and the identification of impurities.

WGMs are produced from travelling waves around the curved dielectric boundary of the cylindrical crystal. They are separated into quasi-transverse magnetic and quasi-transverse electric mode polarisation families, respectively denoted as WGH$_{m,n,p}$ (with the majority of energy stored in the $E_z$, $H_\phi$, and $H_r$ components) and WGE$_{m,n,p}$ (with the majority of energy stored in the $H_z$, $E_\phi$, and $E_r$ components). Mode orders are classified by the positive integer mode numbers $m$, $n$, and $p+\delta$, corresponding to the number of azimuthal, radial, and axial antinodes in the WGM field distribution (where $\delta=1$). The energy storage efficiency that a mode exhibits is represented by its Q-factor, a ratio of the mode frequency to its Full Width at Half Maximum, and this value acts as a benchmark to compare to wider literature results for other dielectric resonators. The fundamental WGMs of each mode family (WGH$_{m,1,1}$ an WGE$_{m,1,1}$) typically exhibit the highest Q and are of key focus in this study. 

The dielectric permittivity tensor of a resonator determines the precise eigenfrequencies of its resonant modes. This TeO$_2$ polymorph exhibits uniaxial anisotropy, splitting the tensor into components parallel ($\varepsilon_{\parallel}$) and perpendicular ($\varepsilon_{\perp}$) to the crystal $c$-axis. The sample in this work is intended for optical birefringence experiments and has been cut with the anisotropy perpendicular to the cylinder $z$-axis, meaning the anisotropy is chosen to be aligned to the $x$-axis and 
\begin{equation}\label{eq:1}
    \varepsilon=\begin{pmatrix}\varepsilon_{\parallel}&0&0\\0&\varepsilon_{\perp}&0\\0&0&\varepsilon_{\perp}\end{pmatrix}=\begin{pmatrix}\varepsilon_{x}&0&0\\0&\varepsilon_{y}&0\\0&0&\varepsilon_{z}\end{pmatrix}.
\end{equation}

\noindent For clarity in this work, all directionally-dependent quantities are expressed in terms of their orientation relative to the crystal $x$-axis (i.e. $\varepsilon_x\equiv\varepsilon_\parallel$, $\varepsilon_y=\varepsilon_z\equiv\varepsilon_\perp$). The direction of the anisotropy is atypical for microwave WGM experiments, with a $z$-cut crystal being standard practice. As a result, resonant modes can separately couple to the crystal lattice and the crystal volume, producing interesting outcomes especially for the WGE mode class. 

To analyse and compare the inherent losses of TeO$_2$, the filling factors $p_i$ of individual modes can be calculated using
\begin{equation}\label{eq:2}
p_i=\frac{\int\int\int_V \varepsilon_i\vec{E}_i\cdot \vec{E}_i^*dv}{\int\int\int_{V_t}\varepsilon(v)\vec{E}\cdot\vec{E}^*dv},
\end{equation}

\noindent To determine the uncertainty in the measured permittivities, the following relation may be implemented:
\begin{equation}\label{eq:2.1}
\frac{\partial f_0}{f_0}
=-\frac{1}{2}\left(
p_{\varepsilon_\parallel}\frac{\partial\varepsilon_\parallel}{\varepsilon_\parallel}
+p_{\varepsilon_\perp}\frac{\partial\varepsilon_\perp}{\varepsilon_\perp}
\right)
-p_D\frac{\partial D}{D}
-p_L\frac{\partial L}{L},
\end{equation}

\noindent where the fractional changes in resonant frequency, $\frac{\partial f_0}{f_0}$, parallel permittivity, $\frac{\partial\varepsilon_\parallel}{\varepsilon_\parallel}$, perpendicular permittivity, $\frac{\partial\varepsilon_\perp}{\varepsilon_\perp}$, crystal diameter, $\frac{\partial D}{D}$, and crystal length, $\frac{\partial L}{L}$, are weighted by the corresponding electric and dimensional filling factors. For whispering-gallery modes, this expression may be further simplified by retaining only the dominant dimensional contribution, associated with the crystal diameter $D$. This allows the uncertainty in $D$ to be related directly to the uncertainty in the relevant permittivity component being determined.

Using the directional filling factors and Q-factors of modes, the frequency and directionally dependent loss tangent tan$\delta_i$ can be determined with
\begin{equation}\label{eq:3}
    Q^{-1}=p_\parallel tan\delta_\parallel+p_\perp tan\delta_\perp+Q_s^{-1}+Q_r^{-1},
\end{equation}

\noindent where the Q at a particular frequency is represented in terms of the sum of the filling factor-weighted loss tangents for each chosen direction. The perpendicular filling factor value is more accurately a sum of the perpendicular directions ($p_\perp\equiv p_{y+z}$). By ignoring negligible surface-resistance and radiation loss terms, $Q_s^{-1}$ and $Q_r^{-1}$, the relevant tan$\delta$ values can be determined simultaneously from two nearby modes.

Introducing a large static magnetic field to the dielectric resonator will cause spin systems within the crystal lattice to become non-degenerate. When the strength of this field is varied, instantaneous energy absorption on resonance will indicate spin system transitions, and considering modes at different frequencies allows the construction of Landé $g$-factors according to the system's spin Hamiltonian:
\begin{equation}\label{eq:4}
    \hat{H}_{spin}=g\mu_B B_0\hat{S}_z+D\hat{S}_z^2+E(\hat{S}_x^2-\hat{S}_y^2)+A\textbf{S}\cdot \textbf{I}.
\end{equation}

\noindent Here, the first term represents the standard electron Zeeman transitions according to the $g$-factor, Bohr magneton $\mu_B$, and magnetic field strength $B_0$. The second and third terms are the axial and rhombic Zero-Field Splitting (ZFS) components according to the constant parameters $D$ and $E$, outlining higher order spin $S\neq\frac{1}{2}$ effects. The final term is the hyperfine component defined by constant parameter $A$, involving the nuclear spin operator $\textbf{I}$ for nuclei with non-zero nuclear spin. Each term in the Hamiltonian involves relevant components of the transition spin operator $\textbf{S}$. Additional negligible or irrelevant terms are discarded \cite{abragam2012electron,GJEdwards_1995}.

\begin{figure}[h]
    \centering
    \includegraphics[width=1\linewidth]{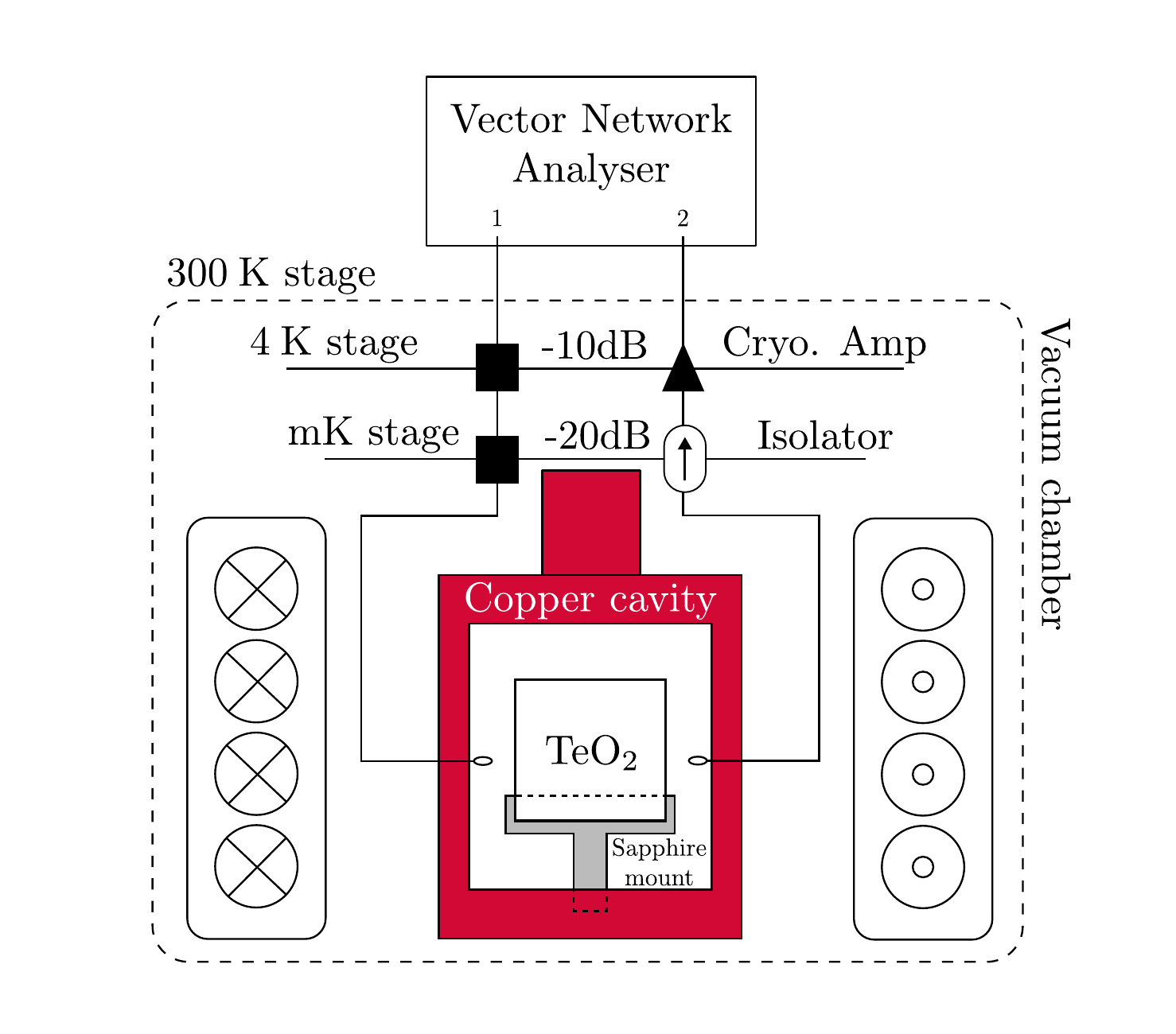}
    \caption{Circuit diagram of the cavity and refrigerator setup, showing the 300 K, 4 K and 20 mK temperature stages. Cryogenic stages are shown to be under a vacuum and within a magnetic field, with loop probes coupling to the field inside the cavity. Probe 1 from the vector network analyser undergoes two stages of attenuation, and probe 2 is connected by an isolator and a 6-20 GHz Low Noise Factory amplifier.}
    \label{fig:1}
\end{figure}

The configuration of this experiment is standard for WGM loss analysis and ESR, and is the same as previously reported \cite{10.1063/5.0224102}. In this setup, pictured in Fig.~\ref{fig:1}, a cylindrical copper resonator cavity is placed in a magnetic field within a dilution refrigerator, with loop probes at the base of the cavity to excite and measure microwave WGMs. The input coaxial cable undergoes two stages of attenuation at the 4 K and mK levels to remove thermal noise, while the output cable has both an isolator and a cryogenic amplifier to prevent feedback noise and amplify the output signal respectively. These cables connect to a vector network analyser for both excitation and S$_{21}$ transmission measurements, allowing for analysis of mode structures and spectroscopy signals. The refrigerator setup produces a minimum measurement temperature of 20 mK, and ESR spectroscopy is performed by sweeping the magnetic field strength up to 1.2 T. 

Uniquely for this work, the TeO$_2$ sample is a cylindrical crystal with no internal or external post structure as part of the geometry. Instead, the sample rests on a machined sapphire mount protruding from the cavity's base as seen in Fig.~\ref{fig:1}. Sapphire's low dielectric loss but good thermal conductivity means that the sample is thermally coupled to the cavity without significant photonic losses.

\section{\label{sec:level3}Mode Characterisation}

To accurately simulate cryogenic WGMs, room temperature characterisation was first performed to match with FEM simulations. This was undertaken by removing the crystal from its cavity to suppress cavity modes, then introducing coaxial loop probes, placing one on a rotating setup. Manual adjustments of the dynamic probe in the azimuthal, axial, and radial directions allow for the identification of modes, as mode numbers are counted from changes in the power spectrum when the probe passes evanescent field near nodes or antinodes. Fundamental modes in both families can then be identified and their eigenfrequencies can be compared to simulations. The FEM software COMSOL was used, constructing the setup using a tetrahedral mesh of $2\times10^5$ elements. The values for the permittivity tensor were adjusted from known literature values to precisely match experimental and simulated modes \cite{PEERCY19751105}.

A selection of high-Q room temperature modes were chosen and tracked across the cooling cycles of the dilution refrigerator to see how the mode eigenfrequencies changed between cryogenic regimes. By adjusting the permittivity tensor to match these new mode solutions and considering the thermal contraction of materials in the cavity, an accurate cryogenic simulation was constructed. Uncertainties for both cryogenic permittivity values were determined with Equation \ref{eq:2.1}, resulting in $\varepsilon_\parallel=25.75\pm0.08$ and $\varepsilon_\perp=20.90\pm0.07$. These were crosschecked against the limited literature for TeO$_2$ at cryogenic temperatures and found to be within error margins.

In similar biaxial crystal studies, the additional degree of freedom in the permittivity tensor is accounted for by considering the doublet effect typically observed at the fundamental mode eigenfrequencies \cite{10.1063/1.4920987,article1}. Due to the considered range for the azimuthal mode number $m$, the mode doublets generally exhibited splitting below 1 MHz and it was appropriate to treat them as degenerate when calculating permittivity.

\begin{figure}[t!]
    \centering
    \includegraphics[width=1\linewidth]{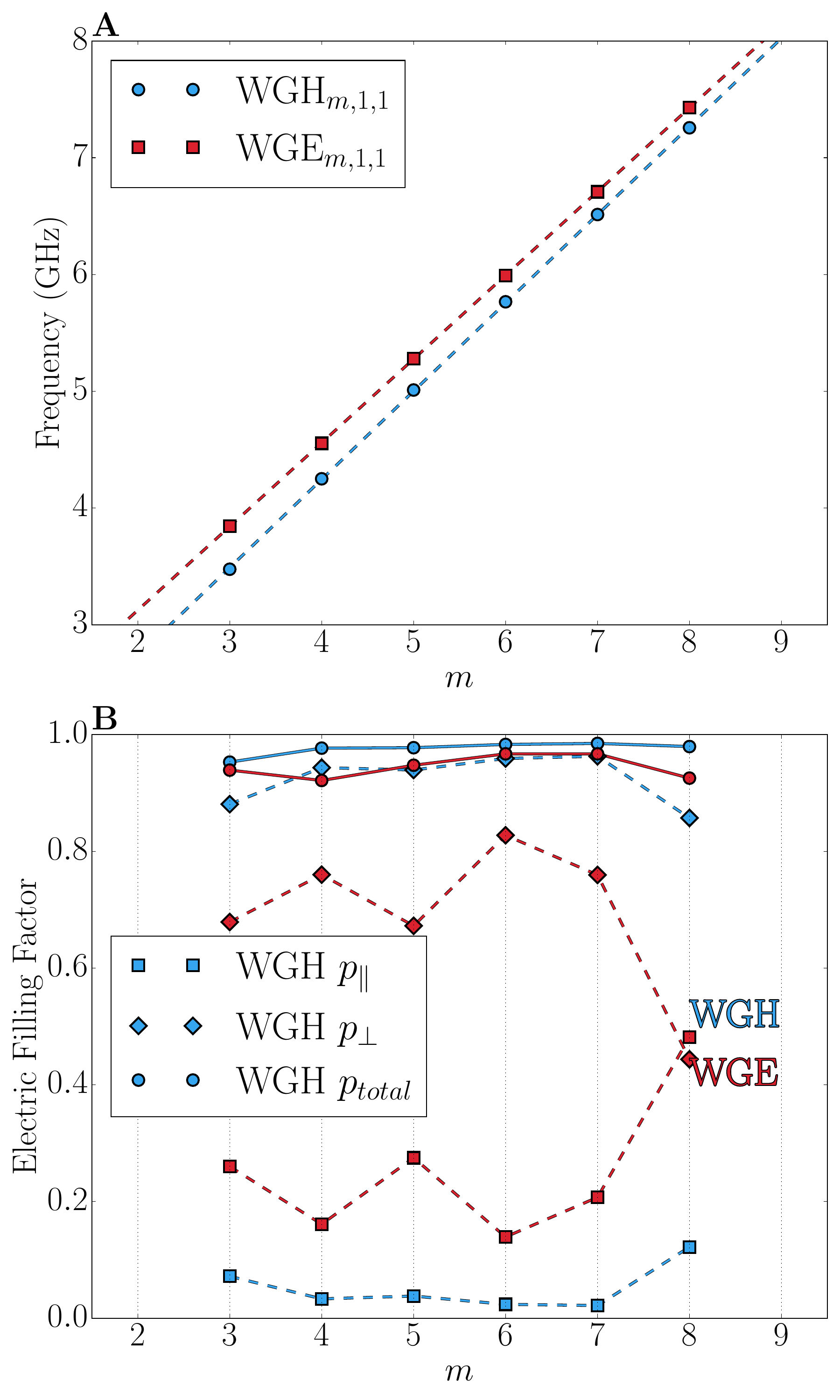}
    \caption{\textbf{A} Simulated mK eigenfrequencies compared to azimuthal mode number $m$ and \textbf{B} directionally dependent electrical filling factors calculated from mK simulations for both WGH (blue) and WGE (red) mode families.}
    \label{fig:4}
\end{figure}

Fig.~\ref{fig:4} (A) shows the cryogenic fundamental WGH$_{m,1,1}$ and WGE$_{m,1,1}$ mode eigenfrequencies against the azimuthal mode number $m$. These were found using visual identification, counting field maxima when mapping $E_z$ or $H_z$ field components. Electrical filling factor calculations were performed within the simulation software, integrating over field values according to Equation \ref{eq:2}. Fig.~\ref{fig:4} (B) contains the calculated filling factors for both mode families, showing both crystallographic directions as well as the total value. 

As the majority of the electric field for a WGH mode is stored in only the $z$-direction, the modes are unaffected by the unusual orientation of the crystal's anisotropy. On the other hand, the WGE modes couple separately to the macroscopic shape of the crystal and its lattice structure, resulting in warped elliptical field distributions for this mode family and non-standard filling factor results. This effect is further discussed in Section \ref{sec:level4}. The filling factors for the WGH modes behave as expected, with the $p_{total}$ value approaching unity as $m$ increases. Additionally, it can be seen that the majority of the energy is found in $p_\perp$,  correlating with the $z$-directional nature of the mode family's electric field. As seen in Fig.~\ref{fig:4} (A), the frequencies of the two mode families approach each other as $m$ increases, which can lead to hybridisation and field mixing losses. This causes a visible change in $p$ for both polarisations.

\section{\label{sec:level4}Losses}

\begin{figure}[t!]
    \centering
    \includegraphics[width=1\linewidth]{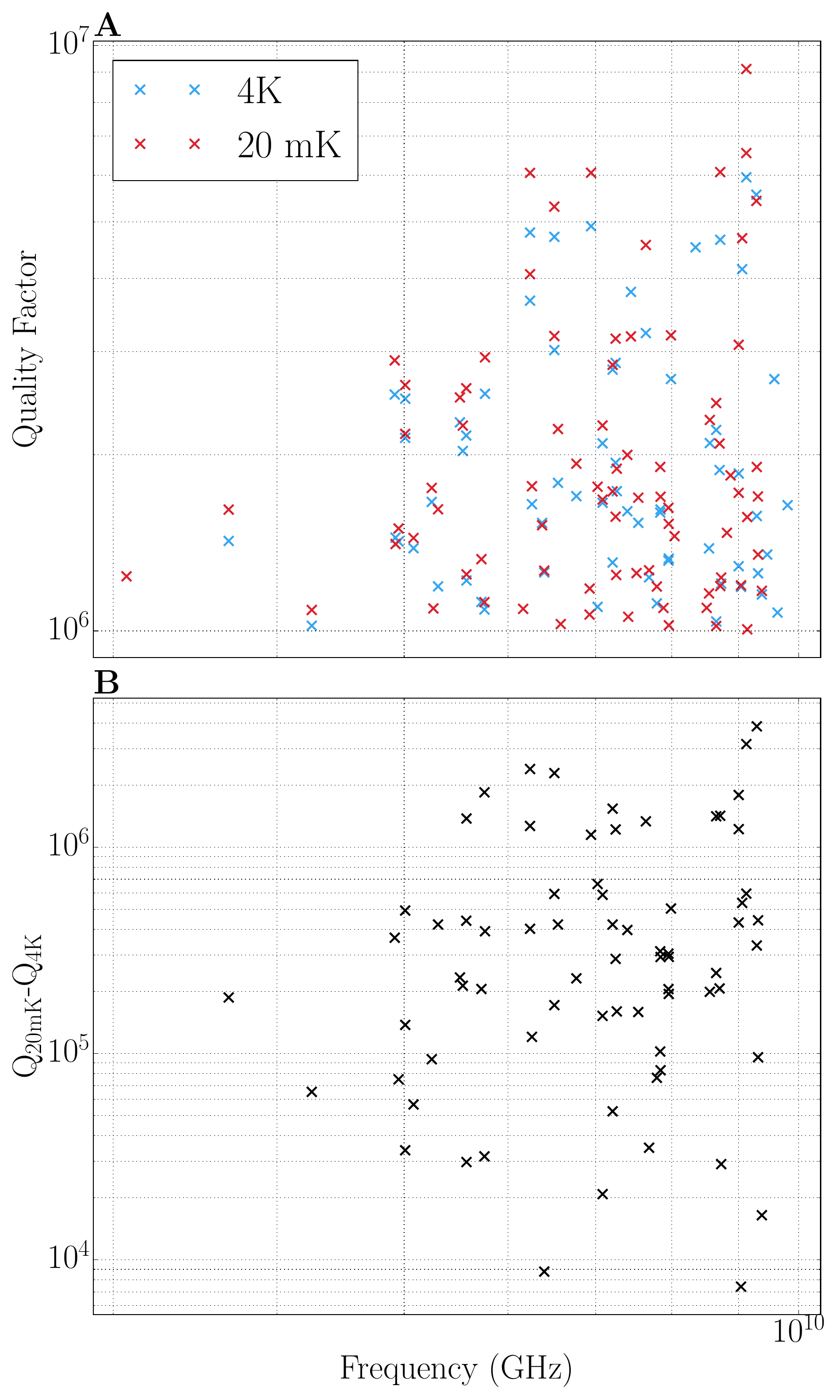}
    \caption{\textbf{A} Quality factor Q against resonant frequency for 20 mK (red) and 4 K (blue) temperature sweeps and \textbf{B} the difference in Quality factor $\mathrm{Q_{20mK}-Q_{4K}}$ between temperatures. Only modes with $Q>10^6$ are shown.}
    \label{fig:2}
\end{figure}

Measurements of WGMs were performed at both 4 K and 20 mK, sweeping the incident power across a range of 8 dBm. Resonant modes within the output power data were fitted to a Lorentzian model and the corresponding Q-factor was determined. Fig.~\ref{fig:2} (A) plots the Qs of the full spectrum of modes irrespective of the family, showing data for both temperatures. 20 mK modes show a consistently higher Q, and the difference in Q between temperatures is plotted in Fig.~\ref{fig:2} (B). 20 mK mode Qs reached up to $9\times10^{6}$ for greater frequencies. Q is also observed to rise significantly as frequency increases. Both trends are expected for modes of this type, as both increasing frequency and lowering temperature remove cavity resistivity contributions to the losses. By further increasing the frequency of the sweeps in future work, a Q ceiling will eventually be reached.

Sapphire remains the benchmark upper limit for WGM Q-factors, with multiple studies confirming values greater than $10^9$ \cite{10.1117/12.2044823,10.1063/1.3595942}. CaWO$_4$ is another high-Q crystal with recent literature interest, exhibiting maximum Q values above $10^7$ \cite{10.1063/5.0224102}. $^{28}$Si is a more common crystal with a range of additional properties, and has supported WGMs with Qs around $10^6$ \cite{dbc7810e9767435d8ab8faba2b66db43,PhysRevApplied.11.044044}. This positions TeO$_2$ favourably, potentially exceeding the quality of CaWO$_4$ resonant modes with further studies at higher frequencies.

\begin{figure}[b!]
    \centering
    \includegraphics[width=1\linewidth]{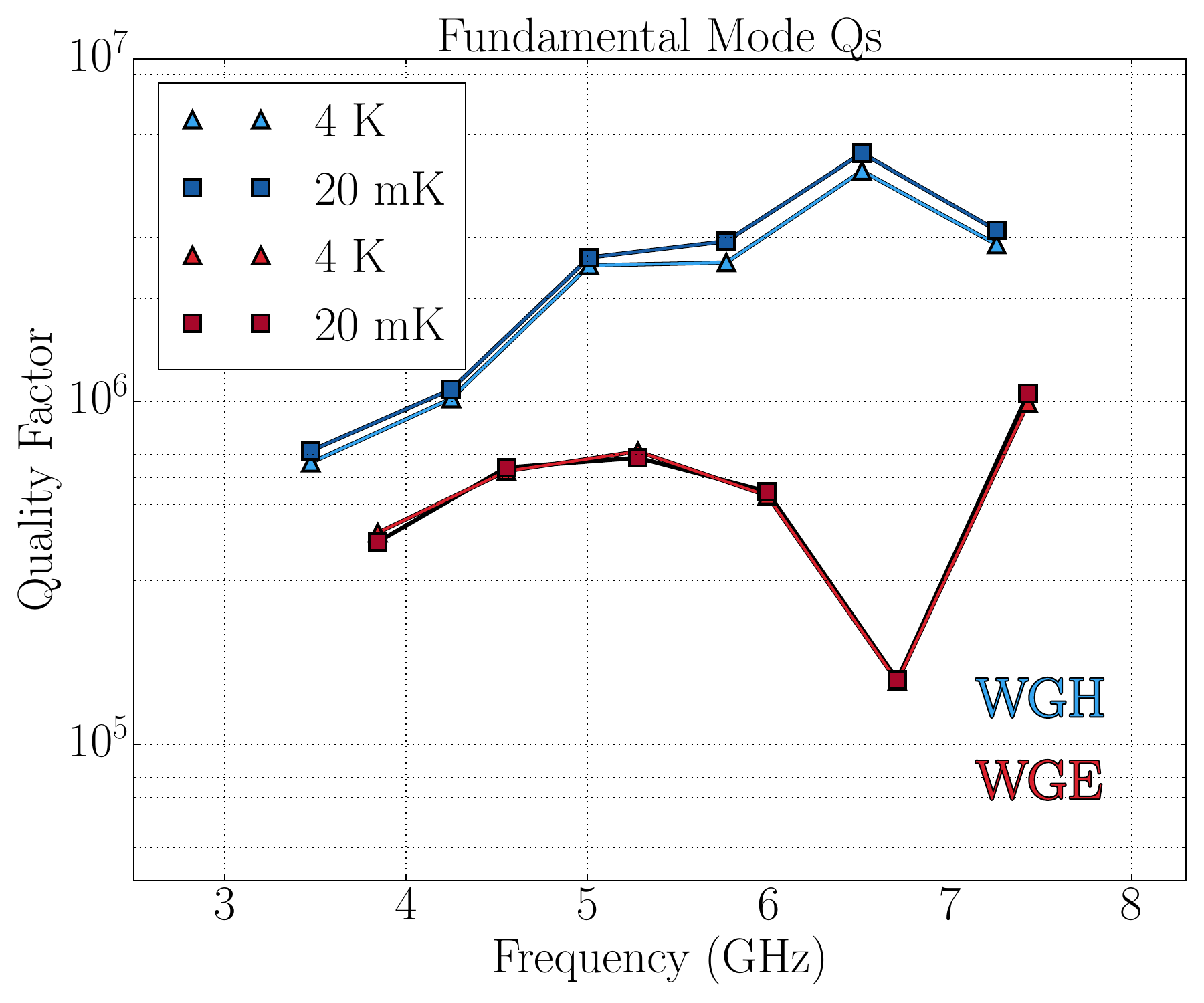}
    \caption{Quality factor Q of fundamental radial/axial WGH$_{m,1,1}$ (blue) and WGE$_{m,1,1}$ (red) modes plotted against their frequency. Both 4 K (triangles) and 20 mK (squares) cases are shown.}
    \label{fig:3}
\end{figure}

Fig.~\ref{fig:3} shows the Q-factors of identified fundamental WGMs for both temperatures and polarisations. This was achieved with the permittivity matched simulations, identifying the fundamental modes from their visual field patterns and comparing eigenfrequencies to resonant peaks on the output power sweep spectra. Similarly to Fig.~\ref{fig:2}, the Qs increase with frequency and decrease at higher temperatures. As seen in filling factor calculations, the crossing of the polarisation families as $m$ increases has led to a decrease in Q from the mixing of field components. Additionally, similar effects regarding WGE polarisation were discussed in Section \ref{sec:level3}, causing inconsistencies in Q as frequency changes. As a result, the WGH modes are the most appropriate to consider on when comparing to wider literature results.

When identifying WGE modes via simulations, the polarisation family appeared to split into two separate groups separated by several hundred MHz. This is the result of the unusual crystal axis direction compared to the crystal shape, causing the standard WGE polarisation to become non-degenerate and form a new family coupled to the crystal geometry rather than the lattice structure itself. These could be visually categorised into the different couplings in simulation stages, as the lattice-coupled family exhibited higher Q and an elliptical wave path. This interesting effect motivates further study, as this is atypical for WGM experiments. By splitting the WGE modes, the filling factor of the volume and crystal lattice are now also unique, and may be considered separately. 

\begin{figure}[h!]
    \centering
    \includegraphics[width=1\linewidth]{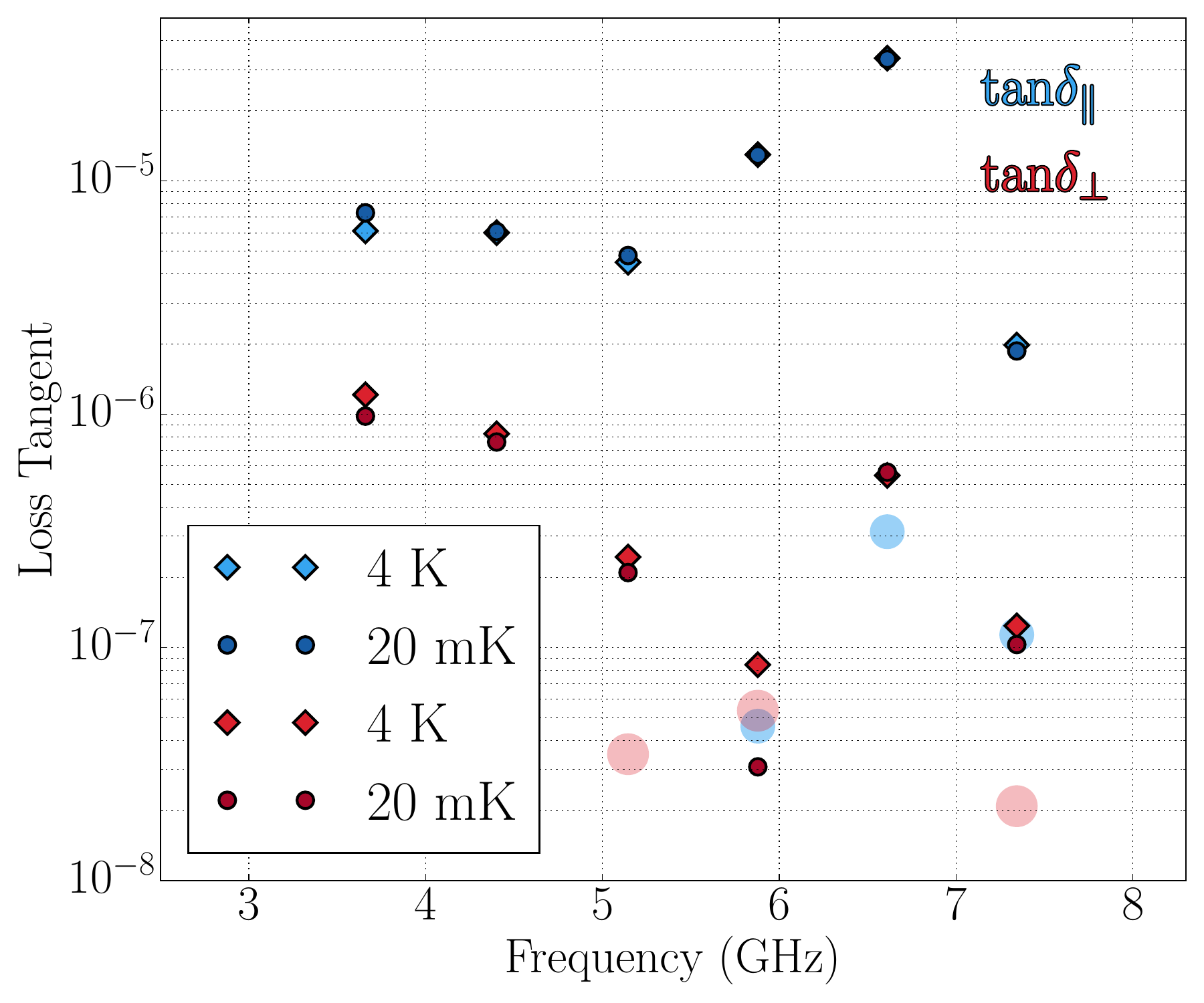}
    \caption{Loss tangent tan$\delta$ values against average mode frequency for both $x$ (blue) and $y-z$ (red) directions, parallel and perpendicular to the crystal axis respectively. Both 4 K (diamonds) and 20 mK (circles) cases are shown.}
    \label{fig:5}
\end{figure}

Data for filling factor and Q were used in Equation \ref{eq:3} to find directional frequency-dependent tan$\delta$ values. Simultaneous calculations were performed with WGH and WGE polarisation pairs at their average frequency. The results of this process are seen in Fig.~\ref{fig:5}. The 20 mK data generally produces a lower tan$\delta$, however the temperatures are comparable. Loss tangent for both directions appears to decrease as frequency increases, reaching as low as tan$\delta_\parallel=2\times10^{-6}$ and tan$\delta_\perp=3\times10^{-8}$. The perpendicular $y-z$ oriented loss tangent is the most reliable when considering the crystal's inherent losses, as it is more closely related to WGH mode data (Using predominantly the $E_z$ electric field). These results once again place TeO$_2$ in good standing with comparable dielectrics, with sapphire reaching tan$\delta$ values below $10^{-9}$, and CaWO$_4$ and $^{28}$Si reaching the order of $10^{-8}$ and $10^{-6}$ respectively \cite{10.1117/12.2044823,10.1063/1.3595942,10.1063/5.0224102,dbc7810e9767435d8ab8faba2b66db43,PhysRevApplied.11.044044}.

The reduction in loss as temperature decreases reveals that TeO$_2$ is resistant to spin impurity effects unlike other crystals in their lowest cryogenic regimes. Sapphire has been found to exhibit two-level-system (TLS) losses from defect participation at single-photon energies, increasing tan$\delta$ as temperature reaches mK levels \cite{10.1063/1.3595942}. The combination of its low loss and the absence of TLS loss positions TeO$_2$ as a promising low-loss dielectric for superconducting microwave or quantum-device applications.

\section{\label{sec:level5}Spectroscopy}

\begin{figure}
    \centering
    \includegraphics[width=1\linewidth]{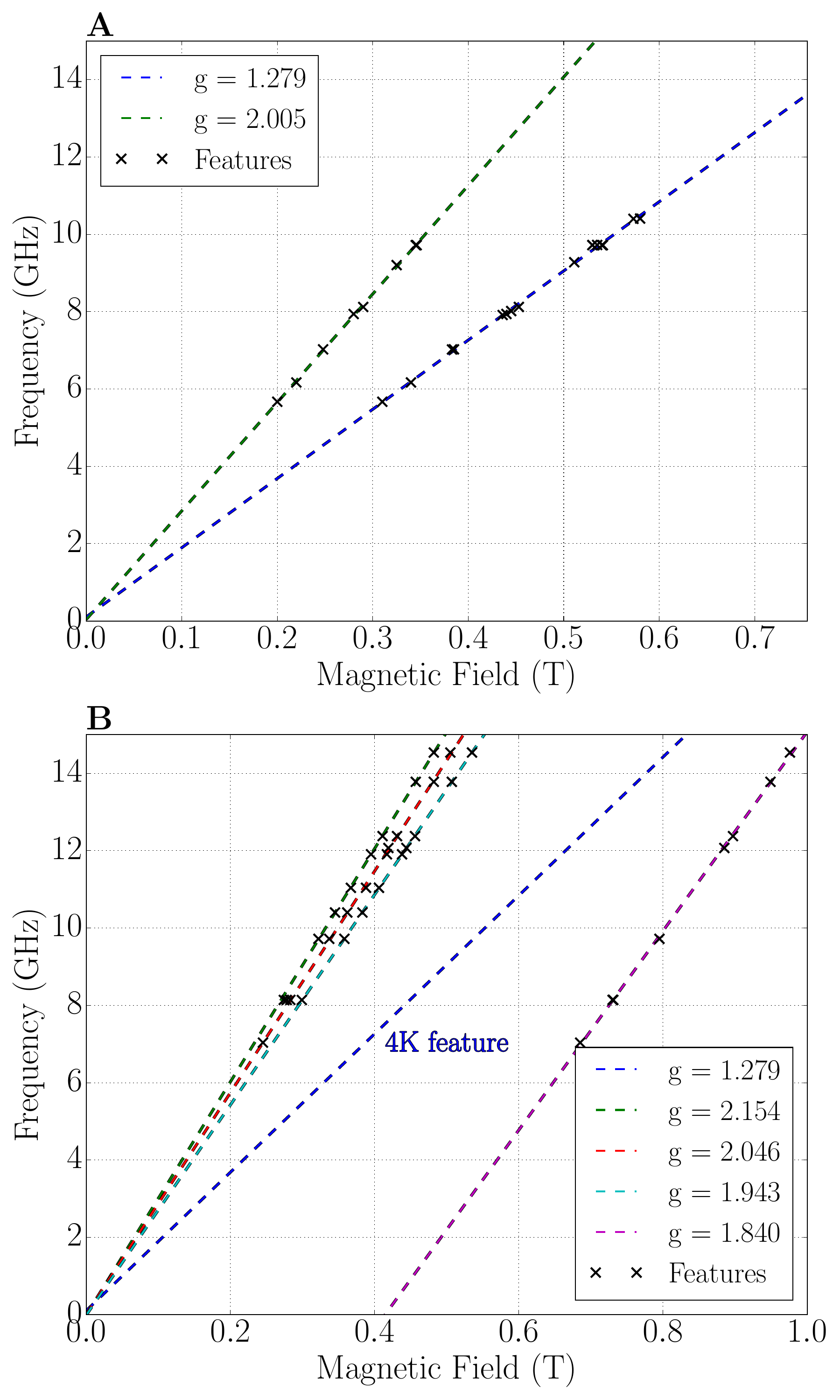}
    \caption{ESR spectroscopy features of resonant modes at \textbf{A} 4 K and \textbf{B} 20 mK plotted against magnetic field strength. $g$-factors are calculated using the magnetic field and frequency of the features, and individual spin systems are represented with differently coloured linear models (blue, green, red, cyan, magenta).}
    \label{fig:7}
\end{figure}

ESR spectroscopy of the TeO$_2$ crystal at 4 K was performed up to 1.2 T for several high-Q WGMs, and significant features in each resonant frequency are plotted in Fig.~\ref{fig:7} (A) according to the magnetic field. Using the linear component of Equation \ref{eq:4} (due to the lack of ZFS), the $g$-factors of two observed features have been identified. The first feature yields $g=2.005$, corresponding with effects that often arise from unpaired electrons at defect sites. TeO$_2$ often contains ionic dopant impurities from transition metals Fe$^{3+}$, V$^{4+}$, Al$^{3+}$, Cr$^{3+}$, and Cr$^{5+}$, as well as oxygen vacancy centres \cite{KRaksanyi_1995,GJEdwards_1995,WATTERICH1986987,WATTERICH1992189,WATTERICH1987249,PhysRevB.32.2533}. Paramagnetic studies of these impurities all observe features nearing $g_e$, suggesting a common ionic impurity is responsible for this effect.

The $g=1.279$ feature indicates a distinct spin system, but the value does not align with any known literature results. This effect may arise from a defect state in the crystal lattice experiencing strong orbital mixing and hence reducing $g$ from the standard $g_e$ value. This is consistent with effects observed in low temperature semiconductors as discussed below.

At 20 mK, several alternate effects were observed. Fig.~\ref{fig:7} (B) presents the significant features, as well as the 4 K $g=1.279$ feature which was notably absent. The $g=2.046$ feature corresponds closely again with $g_e$ however presented as a far broader structure with two notable satellite features reflecting around the central point (at $g=1.943$ and $2.154$).  These effects are not observed in wider established TeO$_2$ paramagnetic studies, suggesting an additional mechanism may be responsible, and further analysis of the broad feature below supports this.

The second fine structure effect observed in the mK regime corresponds with $g=1.840$ and possesses a ZFS of 10.67 GHz. ZFS effects are only observed for $S>\frac{1}{2}$ spin transitions, such as from a metallic dopant site with multiple unpaired parallel electrons. This has been explored in quartz-based experiments with spin systems significantly deviating from $g_e$ \cite{10.1063/1.4858075}.

The $g=1.279$ feature observed at 4 K is absent from the mK results, indicating the `freeze-out' of the spin system. At low temperatures, p and n-type carriers have been observed to vanish in large bandgap semiconductors with comparable properties to TeO$_2$ \cite{PhysRevApplied.21.064002}. The specific effect could be explained by oxygen vacancy sites coupling to local Tellurium atoms, shifting the $g$-factor to the observed value \cite{PhysRevB.32.2533}.

\begin{figure}
    \centering
    \includegraphics[width=1\linewidth]{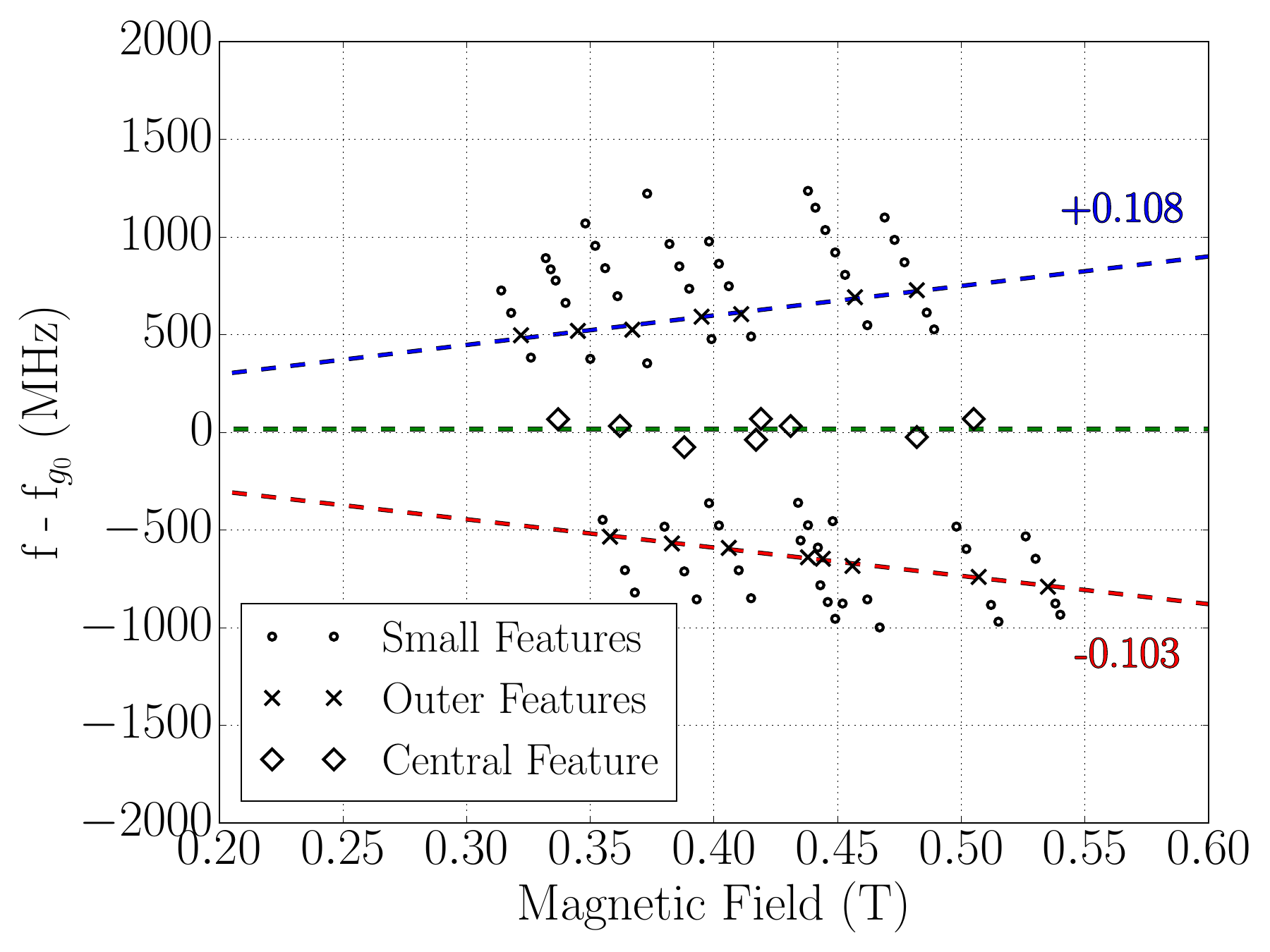}
    \caption{Flattened 20 mK ESR data plotted as the deviation from a central structure at $g_0=2.046$. The central feature (diamonds) is fitted in green along a deviation of 0, and the sharp outer features (crosses) are fitted in blue and red. Additional small features (circles) around the outer features are plotted to demonstrate potential spin system effects.}
    \label{fig:9}
\end{figure}

Further inspection was performed on the broad $g_0=2.046$ shape at 20 mK, with Fig.~\ref{fig:9} displaying all subsidiary points found on the feature flattened around the central peak. The main satellite features appeared as very sharp peaks, linearly diverging by $\Delta g=+0.108,-0.103$ from the centre and exhibiting no ZFS. The chiral structure formed by chains of the TeO$_2$ unit cell is likely the cause of this effect, with local crystal field effects distorting $g$-factors of the same impurity based on its location in the chain along the screw axis.

The most notable features here are the very small peaks running parallel to the outer features. Parallel effects of this size are typical of hyperfine structure transitions caused by non-zero nuclear spin ionic impurities (corresponding to the nuclear spin term in Equation \ref{eq:4}). While the precise quantity of transitions was difficult to observe, it is likely this correlates to a nuclear spin of $I=\frac{5}{2}$ or $\frac{7}{2}$.

\section*{\label{sec:level6}Conclusion}

TeO$_2$ is a promising dielectric candidate for low-loss cryogenic microwave WGM studies, and possesses a range of beneficial properties for fundamental physics experiments and quantum technology applications. Using comparisons between experiment and Finite Element Modelling methods, cryogenic dielectric permittivity values were calculated to be $\varepsilon_\parallel=25.75\pm0.35$ and $\varepsilon_\perp=20.9\pm0.1$.

Using an atypical crystal sample for studies of this type, we have measured maximum WGM Quality factors of $9\times10^6$, with potential for higher values with greater frequencies and samples with standard crystal cut direction. Loss tangent values were calculated from electrical filling factors and determined to reach as low as tan$\delta_\parallel=2\times10^{-6}$ and tan$\delta_\perp=3\times10^{-8}$. Q-factor and tan$\delta$ measurements situate TeO$_2$ well against comparable examples, being comparable to CaWO$_4$, although not as low-loss as sapphire. 

ESR spectroscopy revealed several unique spin systems within the crystal lattice, including a non-zero nuclear spin dopant likely inhabiting sites along the crystal's screw axis and causing diverging $g$-factors. An $S>\frac{1}{2}$ feature was observed with significant ZFS, and a semiconductor carrier freeze-out effect was seen between 4 K and 20 mK spectroscopy measurements. Additional study of several spin systems is required for accurate identification of dopant candidates or other specific effects.

A deep understanding of the losses and spin systems for TeO$_2$ is invaluable for determining its continued use in research and technology. Its low loss suggests its further use in dark matter detection or other microwave resonator applications, and combined with its piezoelectricity and low acoustic loss, the crystal will benefit from increased use in acousto-optical settings. The clean spin environment further positions it for use in quantum technologies requiring a clean spin environment. Further work should be performed on custom samples using a $z$-aligned crystal axis as well as a standard central post support structure within the copper cavity, however additional analysis between geometric and structural filling factors is also of interest. Extending the present experimental work to higher frequencies would additionally allow for a deeper understanding of the crystal's loss characteristics and push it further into the present scientific landscape. 

This work was supported by the Australian Research Council Centre of Excellence for Dark Matter Particle Physics, CE200100008.

\nocite{*}

\bibliography{WGM}

@PREAMBLE{
 "\providecommand{\noopsort}[1]{}" 
 # "\providecommand{\singleletter}[1]{#1}%" 
}

@article{10.1063/5.0224102,
    author = {Hartman, Elrina and Tobar, Michael E. and McAllister, Ben T. and Bourhill, Jeremy and Goryachev, Maxim},
    title = {Precision multi-mode microwave spectroscopy of paramagnetic and rare-earth ion spin defects in single crystal calcium tungstate},
    journal = {Applied Physics Letters},
    volume = {125},
    number = {16},
    pages = {164001},
    year = {2024},
    month = {10},
    doi = {10.1063/5.0224102},
}

@article{ldsb-4dnr,
  title = {Dielectric properties of single-crystal calcium tungstate},
  author = {Hartman, Elrina and Tobar, Michael E. and McAllister, Ben T. and Bourhill, Jeremy and Erb, Andreas and Goryachev, Maxim},
  journal = {Phys. Rev. Appl.},
  volume = {25},
  issue = {3},
  pages = {034055},
  numpages = {9},
  year = {2026},
  month = {Mar},
  publisher = {American Physical Society},
  doi = {10.1103/ldsb-4dnr},
  url = {https://link.aps.org/doi/10.1103/ldsb-4dnr}
}

@article{Creedon2011,
    author = {Creedon, Daniel L. and Reshitnyk, Yarema and Farr, Warrick and Martinis, John M. and Duty, Timothy L. and Tobar, Michael E.},
    title = {High Q-factor sapphire whispering gallery mode microwave resonator at single photon energies and millikelvin temperatures},
    journal = {Applied Physics Letters},
    volume = {98},
    number = {22},
    pages = {222903},
    year = {2011},
    month = {06},
}

@ARTICLE{769347,
  author={Krupka, J. and Derzakowski, K. and Abramowicz, A. and Tobar, M.E. and Geyer, R.G.},
  journal={IEEE Transactions on Microwave Theory and Techniques}, 
  title={Use of whispering-gallery modes for complex permittivity determinations of ultra-low-loss dielectric materials}, 
  year={1999},
  volume={47},
  number={6},
  pages={752-759},
  doi={10.1109/22.769347}}

@article{article2,
author = {Tobar, Michael},
year = {1995},
month = {01},
pages = {276-277},
title = {Gravitational wave detection and low-noise sapphire oscillators},
volume = {12},
journal = {Publications of The Astronomical Society of Australia - PUBL ASTRON SOC AUSTRALIA}
}

@article{PhysRevB.88.224426,
  title = {Ultrasensitive microwave spectroscopy of paramagnetic impurities in sapphire crystals at millikelvin temperatures},
  author = {Farr, Warrick G. and Creedon, Daniel L. and Goryachev, Maxim and Benmessai, Karim and Tobar, Michael E.},
  journal = {Phys. Rev. B},
  volume = {88},
  issue = {22},
  pages = {224426},
  numpages = {8},
  year = {2013},
  month = {Dec},
  publisher = {American Physical Society},
  doi = {10.1103/PhysRevB.88.224426},
  url = {https://link.aps.org/doi/10.1103/PhysRevB.88.224426}
}

@article{dbc7810e9767435d8ab8faba2b66db43,
title = "Determination of low loss in isotopically pure single crystal 28Si at low temperatures and single microwave photon energy",
author = "Nikita Kostylev and Maxim Goryachev and Bulanov, {Andrey D.} and Gavva, {Vladimir A.} and Tobar, {Michael E.}",
year = "2017",
month = mar,
day = "20",
doi = "10.1038/srep44813",
volume = "7",
journal = "Scientific Reports",
issn = "2045-2322",
publisher = "Nature Publishing Group - Macmillan Publishers"
}

@article{10.1063/1.3595942,
    author = {Creedon, Daniel L. and Reshitnyk, Yarema and Farr, Warrick and Martinis, John M. and Duty, Timothy L. and Tobar, Michael E.},
    title = {High Q-factor sapphire whispering gallery mode microwave resonator at single photon energies and millikelvin temperatures},
    journal = {Applied Physics Letters},
    volume = {98},
    number = {22},
    pages = {222903},
    year = {2011},
    month = {06},
    issn = {0003-6951},
    doi = {10.1063/1.3595942},
    url = {https://doi.org/10.1063/1.3595942}
}

@inproceedings{10.1117/12.2044823,
author = {V. S. Ilchenko and A. A. Savchenkov and A. B. Matsko and L. Maleki},
title = {{Crystalline whispering gallery mode resonators: in search of the optimal material}},
volume = {8960},
booktitle = {Laser Resonators, Microresonators, and Beam Control XVI},
editor = {Alexis V. Kudryashov and Alan H. Paxton and Vladimir S. Ilchenko and Lutz Aschke and Kunihiko Washio},
organization = {International Society for Optics and Photonics},
publisher = {SPIE},
pages = {896013},
year = {2014},
doi = {10.1117/12.2044823},
URL = {https://doi.org/10.1117/12.2044823}
}

@article{article3,
author = {Giordano, Vincent and Fluhr, Christophe and Grop, Serge and Dubois, Benoit},
year = {2015},
month = {04},
pages = {},
title = {Tests of Sapphire Crystals Produced with Different Growth Processes for Ultra-stable Microwave Oscillators},
volume = {64},
journal = {IEEE Transactions on Microwave Theory and Techniques},
doi = {10.1109/TMTT.2015.2503748}
}

@article{PhysRevApplied.11.044044,
  title = {Low-Temperature Properties of Whispering-Gallery Modes in Isotopically Pure Silicon-28},
  author = {Bourhill, J. and Goryachev, M. and Creedon, D.L. and Johnson, B.C. and Jamieson, D.N. and Tobar, M.E.},
  journal = {Phys. Rev. Appl.},
  volume = {11},
  issue = {4},
  pages = {044044},
  numpages = {7},
  year = {2019},
  month = {Apr},
  publisher = {American Physical Society},
  doi = {10.1103/PhysRevApplied.11.044044},
  url = {https://link.aps.org/doi/10.1103/PhysRevApplied.11.044044}
}

@article{ARNABOLDI2004775,
title = {CUORE: a cryogenic underground observatory for rare events},
journal = {Nuclear Instruments and Methods in Physics Research Section A: Accelerators, Spectrometers, Detectors and Associated Equipment},
volume = {518},
number = {3},
pages = {775-798},
year = {2004},
issn = {0168-9002},
doi = {https://doi.org/10.1016/j.nima.2003.07.067},
url = {https://www.sciencedirect.com/science/article/pii/S0168900203023374},
author = {C Arnaboldi and F.T {Avignone III} and J Beeman and M Barucci and M Balata and C Brofferio and C Bucci and S Cebrian and R.J Creswick and S Capelli and L Carbone and O Cremonesi and A {de Ward} and E Fiorini and H.A Farach and G Frossati and A Giuliani and D Giugni and P Gorla and E.E Haller and I.G Irastorza and R.J McDonald and A Morales and E.B Norman and P Negri and A Nucciotti and M Pedretti and C Pobes and V Palmieri and M Pavan and G Pessina and S Pirro and E Previtali and C Rosenfeld and A.R Smith and M Sisti and G Ventura and M Vanzini and L Zanotti}
}

@article{Avignone_2024,
doi = {10.1088/1748-0221/19/09/P09013},
url = {https://dx.doi.org/10.1088/1748-0221/19/09/P09013},
year = {2024},
month = {sep},
publisher = {IOP Publishing},
volume = {19},
number = {09},
pages = {P09013},
author = {Avignone, F.T. and Barabash, A.S. and Berest, V. and Bergé, L. and Calvo-Mozota, J.M. and Carniti, P. and Chapellier, M. and Dafinei, I. and Danevich, F.A. and Dumoulin, L. and Ferella, F. and Ferri, F. and Gallas, A. and Giuliani, A. and Gotti, C. and Gras, P. and Ianni, A. and Imbert, L. and Khalife, H. and Kobychev, V.V. and Konovalov, S.I. and Loaiza, P. and de Marcillac, P. and Marnieros, S. and Marrache-Kikuchi, C.A. and Martinez, M. and Nisi, S. and Nones, C. and Olivieri, E. and Ortiz de Solórzano, A. and Peinaud, Y. and Pessina, G. and Poda, D.V. and Rosier, Ph. and Scarpaci, J.A. and Tretyak, V.I. and Umatov, V.I. and Zarytskyy, M.M. and Zolotarova, A.},
title = {Development of large-volume 130TeO2 bolometers for the CROSS 2{$\beta$} decay search experiment},
journal = {Journal of Instrumentation}
}

@article{LCardani_2012,
doi = {10.1088/1748-0221/7/01/P01020},
url = {https://dx.doi.org/10.1088/1748-0221/7/01/P01020},
year = {2012},
month = {jan},
publisher = {},
volume = {7},
number = {01},
pages = {P01020},
author = {L Cardani and L Gironi and J W Beeman and I Dafinei and Z Ge and G Pessina and S Pirro and Y Zhu},
title = {Performance of a large TeO2 crystal as a cryogenic bolometer in searching for neutrinoless double beta decay},
journal = {Journal of Instrumentation}
}

@article{article,
author = {Bandac, Iulian and Barabash, A. and Bergé, L. and Brière, M. and Bourgeois, Chrystal and Carniti, Paolo and Chapellier, Maurice and Combarieu, M. and Dafinei, Ioan and Danevich, Fedor and Dosme, N. and Doullet, D. and Dumoulin, L. and Ferri, F. and Giuliani, A. and Gotti, C. and Gras, P. and Guerard, E. and Ianni, Aldo and Zolotarova, A.},
year = {2020},
month = {01},
pages = {},
title = {The 0{$\nu$}2{$\beta$}-decay CROSS experiment: preliminary results and prospects},
volume = {2020},
journal = {Journal of High Energy Physics},
doi = {10.1007/JHEP01(2020)018}
}

@article{Nagel2025,
	author = {Nagel, Moritz and Parker, Stephen R. and Kovalchuk, Evgeny V. and Stanwix, Paul L. and Hartnett, John G. and Ivanov, Eugene N. and Peters, Achim and Tobar, Michael E.},
	journal = {Nature Communications},
	number = {1},
	pages = {8174},
	title = {Direct terrestrial test of Lorentz symmetry in electrodynamics to 10-18},
	volume = {6},
	year = {2015}}

@article{Tobar2010,
  title = {Testing local Lorentz and position invariance and variation of fundamental constants by searching the derivative of the comparison frequency between a cryogenic sapphire oscillator and hydrogen maser},
  author = {Tobar, Michael Edmund and Wolf, Peter and Bize, S\'ebastien and Santarelli, Giorgio and Flambaum, Victor},
  journal = {Phys. Rev. D},
  volume = {81},
  issue = {2},
  pages = {022003},
  numpages = {10},
  year = {2010},
  month = {Jan},
  publisher = {American Physical Society},
  doi = {10.1103/PhysRevD.81.022003},
  url = {https://link.aps.org/doi/10.1103/PhysRevD.81.022003}
}

@article{Tobar2013,
  title = {Testing local position and fundamental constant invariance due to periodic gravitational and boost using long-term comparison of the SYRTE atomic fountains and H-masers},
  author = {Tobar, M. E. and Stanwix, P. L. and McFerran, J. J. and Gu\'ena, J. and Abgrall, M. and Bize, S. and Clairon, A. and Laurent, Ph. and Rosenbusch, P. and Rovera, D. and Santarelli, G.},
  journal = {Phys. Rev. D},
  volume = {87},
  issue = {12},
  pages = {122004},
  numpages = {8},
  year = {2013},
  month = {Jun},
  publisher = {American Physical Society},
  doi = {10.1103/PhysRevD.87.122004},
  url = {https://link.aps.org/doi/10.1103/PhysRevD.87.122004}
}

@article{Bushev2019,
  title = {Testing the generalized uncertainty principle with macroscopic mechanical oscillators and pendulums},
  author = {Bushev, P. A. and Bourhill, J. and Goryachev, M. and Kukharchyk, N. and Ivanov, E. and Galliou, S. and Tobar, M. E. and Danilishin, S.},
  journal = {Phys. Rev. D},
  volume = {100},
  issue = {6},
  pages = {066020},
  numpages = {7},
  year = {2019},
  month = {Sep},
  publisher = {American Physical Society},
  doi = {10.1103/PhysRevD.100.066020},
  url = {https://link.aps.org/doi/10.1103/PhysRevD.100.066020}
}

@article{SCHAFFNER201530,
title = {Particle discrimination in TeO2 bolometers using light detectors read out by transition edge sensors},
journal = {Astroparticle Physics},
volume = {69},
pages = {30-36},
year = {2015},
issn = {0927-6505},
doi = {https://doi.org/10.1016/j.astropartphys.2015.03.008},
url = {https://www.sciencedirect.com/science/article/pii/S0927650515000481},
author = {K. Schäffner and G. Angloher and F. Bellini and N. Casali and F. Ferroni and D. Hauff and S.S. Nagorny and L. Pattavina and F. Petricca and S. Pirro and F. Pröbst and F. Reindl and W. Seidel and R. Strauss}
}

@article{ARNABOLDI20102999,
title = {Production of high purity TeO2 single crystals for the study of neutrinoless double beta decay},
journal = {Journal of Crystal Growth},
volume = {312},
number = {20},
pages = {2999-3008},
year = {2010},
issn = {0022-0248},
doi = {https://doi.org/10.1016/j.jcrysgro.2010.06.034},
url = {https://www.sciencedirect.com/science/article/pii/S0022024810004343},
author = {C. Arnaboldi and C. Brofferio and A. Bryant and C. Bucci and L. Canonica and S. Capelli and M. Carrettoni and M. Clemenza and I. Dafinei and S. {Di Domizio} and F. Ferroni and E. Fiorini and Z. Ge and A. Giachero and L. Gironi and A. Giuliani and P. Gorla and E. Guardincerri and R. Kadel and K. Kazkaz and L. Kogler and Y. Kolomensky and J. Larsen and M. Laubenstein and Y. Li and C. Maiano and M. Martinez and R. Maruyama and S. Nisi and C. Nones and Eric B. Norman and A. Nucciotti and F. Orio and L. Pattavina and M. Pavan and G. Pessina and S. Pirro and E. Previtali and C. Rusconi and Nicholas D. Scielzo and M. Sisti and Alan R. Smith and W. Tian and M. Vignati and H. Wang and Y. Zhu}
}

@article{10.1063/1.5031485,
    author = {Brofferio, Chiara and Dell’Oro, Stefano},
    title = {Contributed Review: The saga of neutrinoless double beta decay search with TeO2 thermal detectors},
    journal = {Review of Scientific Instruments},
    volume = {89},
    number = {12},
    pages = {121502},
    year = {2018},
    month = {12},
    issn = {0034-6748},
    doi = {10.1063/1.5031485},
    url = {https://doi.org/10.1063/1.5031485}
}

@article{CASALI201744,
title = {Model for the Cherenkov light emission of TeO2 cryogenic calorimeters},
journal = {Astroparticle Physics},
volume = {91},
pages = {44-50},
year = {2017},
issn = {0927-6505},
doi = {https://doi.org/10.1016/j.astropartphys.2017.03.004},
url = {https://www.sciencedirect.com/science/article/pii/S0927650517300828},
author = {N. Casali}
}

@article{Shao:19,
author = {Linbo Shao and Mengjie Yu and Smarak Maity and Neil Sinclair and Lu Zheng and Cleaven Chia and Amirhassan Shams-Ansari and Cheng Wang and Mian Zhang and Keji Lai and Marko Lon\v{c}ar},
journal = {Optica},
number = {12},
pages = {1498--1505},
publisher = {Optica Publishing Group},
title = {Microwave-to-optical conversion using lithium niobate thin-film acoustic resonators},
volume = {6},
month = {Dec},
year = {2019},
url = {https://opg.optica.org/optica/abstract.cfm?URI=optica-6-12-1498},
doi = {10.1364/OPTICA.6.001498},
}

@article{
doi:10.1126/science.abj4396,
author = {Andreas Boes  and Lin Chang  and Carsten Langrock  and Mengjie Yu  and Mian Zhang  and Qiang Lin  and Marko Lončar  and Martin Fejer  and John Bowers  and Arnan Mitchell },
title = {Lithium niobate photonics: Unlocking the electromagnetic spectrum},
journal = {Science},
volume = {379},
number = {6627},
pages = {eabj4396},
year = {2023},
doi = {10.1126/science.abj4396},
URL = {https://www.science.org/doi/abs/10.1126/science.abj4396},
eprint = {https://www.science.org/doi/pdf/10.1126/science.abj4396}
}

@article{10.1063/1.1657275,
    author = {Uchida, Naoya and Ohmachi, Yoshiro},
    title = {Elastic and Photoelastic Properties of TeO2 Single Crystal},
    journal = {Journal of Applied Physics},
    volume = {40},
    number = {12},
    pages = {4692-4695},
    year = {1969},
    month = {11},
    issn = {0021-8979},
    doi = {10.1063/1.1657275},
    url = {https://doi.org/10.1063/1.1657275}
}

@article{10.1121/1.405652,
    author = {Troedson, Shaun C. and Lindsay, Anthony C. and Fuss, Ian G.},
    title = {Nonlinear acoustic phenomena in TeO2},
    journal = {The Journal of the Acoustical Society of America},
    volume = {93},
    number = {1},
    pages = {148-153},
    year = {1993},
    month = {01},
    issn = {0001-4966},
    doi = {10.1121/1.405652},
    url = {https://doi.org/10.1121/1.405652}
}

@article{MOUFOK2019102315,
title = {Electronic structure and optical properties of TeO2 polymorphs},
journal = {Results in Physics},
volume = {13},
pages = {102315},
year = {2019},
issn = {2211-3797},
doi = {https://doi.org/10.1016/j.rinp.2019.102315},
url = {https://www.sciencedirect.com/science/article/pii/S2211379717306952},
author = {Samira Moufok and Lamine Kadi and Bouhalouane Amrani and Kouider Driss Khodja}
}

@article{10.1063/1.3406135,
    author = {Li, Yanlu and Fan, Weiliu and Sun, Honggang and Cheng, Xiufeng and Li, Pan and Zhao, Xian},
    title = {Structural, electronic, and optical properties of {$\alpha$}, {$\beta$}, and {$\gamma$}-TeO2},
    journal = {Journal of Applied Physics},
    volume = {107},
    number = {9},
    pages = {093506},
    year = {2010},
    month = {05},
    issn = {0021-8979},
    doi = {10.1063/1.3406135},
    url = {https://doi.org/10.1063/1.3406135}
}

@article{PhysRevApplied.22.044065,
  title = {Native defects and $p$-type dopability in transparent $\ensuremath{\beta}$-${\mathrm{Te}\mathrm{O}}_{2}$: A first-principles study},
  author = {Huyen, Vu Thi Ngoc and Bae, Soungmin and Costa-Amaral, Rafael and Kumagai, Yu},
  journal = {Phys. Rev. Appl.},
  volume = {22},
  issue = {4},
  pages = {044065},
  numpages = {10},
  year = {2024},
  month = {Oct},
  publisher = {American Physical Society},
  doi = {10.1103/PhysRevApplied.22.044065},
  url = {https://link.aps.org/doi/10.1103/PhysRevApplied.22.044065}
}

@article{https://doi.org/10.1002/pssr.202300271,
author = {Keerthana and Venimadhav, Adyam},
title = {TeO2: A Prospective High-k Dielectric},
journal = {physica status solidi (RRL) – Rapid Research Letters},
volume = {18},
number = {2},
pages = {2300271},
doi = {https://doi.org/10.1002/pssr.202300271},
url = {https://onlinelibrary.wiley.com/doi/abs/10.1002/pssr.202300271},
eprint = {https://onlinelibrary.wiley.com/doi/pdf/10.1002/pssr.202300271},
year = {2024}
}

@article{PAThomas_1988,
doi = {10.1088/0022-3719/21/25/009},
url = {https://dx.doi.org/10.1088/0022-3719/21/25/009},
year = {1988},
month = {sep},
publisher = {},
volume = {21},
number = {25},
pages = {4611},
author = {P A Thomas},
title = {The crystal structure and absolute optical chirality of paratellurite, {$\alpha$}-TeO2},
journal = {Journal of Physics C: Solid State Physics}
}

@article{PhysRevB.4.3736,
  title = {Optical Properties of Single-Crystal Paratellurite (Te${\mathrm{O}}_{2}$)},
  author = {Uchida, Naoya},
  journal = {Phys. Rev. B},
  volume = {4},
  issue = {10},
  pages = {3736--3745},
  numpages = {0},
  year = {1971},
  month = {Nov},
  publisher = {American Physical Society},
  doi = {10.1103/PhysRevB.4.3736},
  url = {https://link.aps.org/doi/10.1103/PhysRevB.4.3736}
}

@article{PEERCY19751105,
title = {Temperature and pressure dependences of the properties and phase transition in paratellurite (TeO2: Ultrasonic, dielectric and Raman and Brillouin scattering results},
journal = {Journal of Physics and Chemistry of Solids},
volume = {36},
number = {10},
pages = {1105-1122},
year = {1975},
issn = {0022-3697},
doi = {https://doi.org/10.1016/0022-3697(75)90053-0},
url = {https://www.sciencedirect.com/science/article/pii/0022369775900530},
author = {P.S. Peercy and I.J. Fritz and G.A. Samara}
}

@Article{TF9494500155,
author ="Szigeti, B.",
title  ="Polarisability and dielectric constant of ionic crystals",
journal  ="Trans. Faraday Soc. ",
year  ="1949",
volume  ="45",
issue  ="0",
pages  ="155-166",
publisher  ="The Royal Society of Chemistry",
doi  ="10.1039/TF9494500155",
url  ="http://dx.doi.org/10.1039/TF9494500155"}

@article{KRaksanyi_1995,
doi = {10.1088/0953-8984/7/14/025},
url = {https://dx.doi.org/10.1088/0953-8984/7/14/025},
year = {1995},
month = {apr},
publisher = {},
volume = {7},
number = {14},
pages = {2889},
author = {K Raksanyi and A Watterich and O R Gilliam and L A Kappers and G J Edwards},
title = {Electron spin resonance of Fe3+ centres in alpha -TeO2:Fe},
journal = {Journal of Physics: Condensed Matter}
}

@article{GJEdwards_1995,
doi = {10.1088/0953-8984/7/15/008},
url = {https://dx.doi.org/10.1088/0953-8984/7/15/008},
year = {1995},
month = {apr},
publisher = {},
volume = {7},
number = {15},
pages = {3013},
author = {G J Edwards and O R Gilliam and R H Bartram and A Watterich and R Voszka and J R Niklas and S Greulich-Weber and J -M Spaeth},
title = {An electron spin resonance study of vanadium-doped alpha -TeO2 single crystals},
journal = {Journal of Physics: Condensed Matter}
}

@article{PhysRevB.32.2533,
  title = {ESR identification of radiation-induced oxygen vacancy centers in paratellurite},
  author = {Watterich, A. and Bartram, R. H. and Gilliam, O. R. and Kappers, L. A. and Edwards, G. J. and F\"oldv\'ari, I. and Voszka, R.},
  journal = {Phys. Rev. B},
  volume = {32},
  issue = {4},
  pages = {2533--2537},
  numpages = {0},
  year = {1985},
  month = {Aug},
  publisher = {American Physical Society},
  doi = {10.1103/PhysRevB.32.2533},
  url = {https://link.aps.org/doi/10.1103/PhysRevB.32.2533}
}

@article{WATTERICH1986987,
title = {Electron spin resonance of aluminum-related color centers in {$\alpha$}-TeO2:Al},
journal = {Journal of Physics and Chemistry of Solids},
volume = {47},
number = {10},
pages = {987-991},
year = {1986},
issn = {0022-3697},
doi = {https://doi.org/10.1016/0022-3697(86)90113-7},
url = {https://www.sciencedirect.com/science/article/pii/0022369786901137},
author = {A. Watterich and R.H. Bartram and O.R. Gilliam and L.A. Kappers and G.J. Edwards and R. Voszka and I. Cravero}
}

@article{WATTERICH1987249,
title = {Electron spin resonance of Cr5+ in TeO2:Cr},
journal = {Journal of Physics and Chemistry of Solids},
volume = {48},
number = {3},
pages = {249-253},
year = {1987},
issn = {0022-3697},
doi = {https://doi.org/10.1016/0022-3697(87)90020-5},
url = {https://www.sciencedirect.com/science/article/pii/0022369787900205},
author = {A. Watterich and R.H. Bartram and G.J. Edwards and O.R. Gilliam and I. Földvári and R. Voszka}
}

@article{WATTERICH1992189,
title = {Electron spin resonance of Cr3+ and perturbed Cr3+ centers in {$\alpha$}-TeO2:Cr},
journal = {Journal of Physics and Chemistry of Solids},
volume = {53},
number = {1},
pages = {189-195},
year = {1992},
issn = {0022-3697},
doi = {https://doi.org/10.1016/0022-3697(92)90027-B},
url = {https://www.sciencedirect.com/science/article/pii/002236979290027B},
author = {A. Watterich and K. Raksányi and O.R. Gilliam and R.H. Bartram and L.A. Kappers and H. Söthe and J.-M. Spaeth}
}

@phdthesis{6152a004901e4d9aaac8424819a98d44,
title = "A 50 K dual-mode sapphire oscillator and whispering spherical mode oscillators",
author = "James Anstie",
year = "2007"
}

@article{Alessandria_2013,
doi = {10.1088/1475-7516/2013/01/038},
url = {https://dx.doi.org/10.1088/1475-7516/2013/01/038},
year = {2013},
month = {jan},
publisher = {},
volume = {2013},
number = {01},
pages = {038},
author = {F. Alessandria and R. Ardito and D.R. Artusa and F.T. Avignone III and O. Azzolini and M. Balata and T.I. Banks and G. Bari and J. Beeman and F. Bellini and A. Bersani and M. Biassoni and T. Bloxham and C. Brofferio and C. Bucci and X.Z. Cai and L. Canonica and S. Capelli and L. Carbone and L. Cardani and M. Carrettoni and N. Casali and N. Chott and M. Clemenza and C. Cosmelli and O. Cremonesi and R.J. Creswick and I. Dafinei and A. Dally and V. Datskov and A. De Biasi and M.P. Decowski and M.M. Deninno and S. Di Domizio and M.L. di Vacri and L. Ejzak and R. Faccini and D.Q. Fang and H.A. Farach and E. Ferri and F. Ferroni and E. Fiorini and M.A. Franceschi and S.J. Freedman and B.K. Fujikawa and A. Giachero and L. Gironi and A. Giuliani and J. Goett and P. Gorla and C. Gotti and E. Guardincerri and T.D. Gutierrez and E.E. Haller and K. Han and K.M. Heeger and H.Z. Huang and R. Kadel and K. Kazkaz and G. Keppel and L. Kogler and Yu. G. Kolomensky and D. Lenz and Y.L. Li and C. Ligi and X. Liu and Y.G. Ma and C. Maiano and M. Maino and M. Martinez and R.H. Maruyama and N. Moggi and S. Morganti and T. Napolitano and S. Newman and S. Nisi and C. Nones and E.B. Norman and A. Nucciotti and F. Orio and D. Orlandi and J.L. Ouellet and M. Pallavicini and V. Palmieri and L. Pattavina and M. Pavan and M. Pedretti and G. Pessina and S. Pirro and E. Previtali and V. Rampazzo and F. Rimondi and C. Rosenfeld and C. Rusconi and S. Sangiorgio and N.D. Scielzo and M. Sisti and A.R. Smith and F. Stivanello and L. Taffarello and M. Tenconi and W.D. Tian and C. Tomei and S. Trentalange and G. Ventura and M. Vignati and B.S. Wang and H.W. Wang and C.A. Whitten Jr. and T. Wise and A. Woodcraft and L. Zanotti and C. Zarra and B.X. Zhu and S. Zucchelli},
title = {The low energy spectrum of TeO2 bolometers: results and dark matter perspectives for the CUORE-0 and CUORE experiments},
journal = {Journal of Cosmology and Astroparticle Physics}
}

@INPROCEEDINGS{4319236,
  author={Hartnett, John G. and Locke, Clayton R. and Ivanov, Eugene N. and Tobar, Michael E. and Stanwix, Paul L.},
  booktitle={2007 IEEE International Frequency Control Symposium Joint with the 21st European Frequency and Time Forum}, 
  title={Cryogenic sapphire oscillator with exceptionally high long-term frequency stability}, 
  year={2007},
  volume={},
  number={},
  pages={1028-1031},
  doi={10.1109/FREQ.2007.4319236}}

@article{Chiappina:23,
author = {Piero Chiappina and Jash Banker and Srujan Meesala and David Lake and Steven Wood and Oskar Painter},
journal = {Opt. Express},
number = {14},
pages = {22914--22927},
publisher = {Optica Publishing Group},
title = {Design of an ultra-low mode volume piezo-optomechanical quantum transducer},
volume = {31},
month = {Jul},
year = {2023},
url = {https://opg.optica.org/oe/abstract.cfm?URI=oe-31-14-22914},
doi = {10.1364/OE.493532},
}

@article{10.1063/1.1805717,
    author = {Ledbetter, Hassel and Leisure, Robert G. and Migliori, Albert and Betts, Jon and Ogi, Hirotsugu},
    title = {Low-temperature elastic and piezoelectric constants of paratellurite ({$\alpha$}-TeO2)},
    journal = {Journal of Applied Physics},
    volume = {96},
    number = {11},
    pages = {6201-6206},
    year = {2004},
    month = {12},
    issn = {0021-8979},
    doi = {10.1063/1.1805717},
    url = {https://doi.org/10.1063/1.1805717}
}

@article{10.1063/1.1659223,
    author = {Ohmachi, Yoshiro and Uchida, Naoya},
    title = {Temperature Dependence of Elastic, Dielectric, and Piezoelectric Constants in TeO2 Single Crystals},
    journal = {Journal of Applied Physics},
    volume = {41},
    number = {6},
    pages = {2307-2311},
    year = {1970},
    month = {05},
    issn = {0021-8979},
    doi = {10.1063/1.1659223},
    url = {https://doi.org/10.1063/1.1659223}
}

@article{PhysRevLett.115.013601,
  title = {Observation of Photon Echoes From Evanescently Coupled Rare-Earth Ions in a Planar Waveguide},
  author = {Marzban, Sara and Bartholomew, John G. and Madden, Stephen and Vu, Khu and Sellars, Matthew J.},
  journal = {Phys. Rev. Lett.},
  volume = {115},
  issue = {1},
  pages = {013601},
  numpages = {5},
  year = {2015},
  month = {Jul},
  publisher = {American Physical Society},
  doi = {10.1103/PhysRevLett.115.013601},
  url = {https://link.aps.org/doi/10.1103/PhysRevLett.115.013601}
}

@article{article7,
author = {Alghadeer, Mohammed and Banerjee, Archan and Lee, Kyunghoon and Hussein, Hussein and Fariborzi, Hossein and Rao, Saleem},
year = {2024},
month = {11},
pages = {},
title = {Mitigating coherent loss in superconducting circuits using molecular self-assembled monolayers},
volume = {14},
journal = {Scientific Reports},
doi = {10.1038/s41598-024-77227-7}
}

@article{MAZZOCCHI20191,
title = {99.992$\%$ 28Si CVD-grown epilayer on 300mm substrates for large scale integration of silicon spin qubits},
journal = {Journal of Crystal Growth},
volume = {509},
pages = {1-7},
year = {2019},
issn = {0022-0248},
doi = {https://doi.org/10.1016/j.jcrysgro.2018.12.010},
url = {https://www.sciencedirect.com/science/article/pii/S0022024818306225},
author = {V. Mazzocchi and P.G. Sennikov and A.D. Bulanov and M.F. Churbanov and B. Bertrand and L. Hutin and J.P. Barnes and M.N. Drozdov and J.M. Hartmann and M. Sanquer}
}

@book{abragam2012electron,
  title={Electron Paramagnetic Resonance of Transition Ions},
  author={Abragam, A. and Bleaney, B.},
  isbn={9780191023002},
  series={Oxford Classic Texts in the Physical Sciences},
  url={https://books.google.com.au/books?id=ASNoAgAAQBAJ},
  year={2012},
  publisher={OUP Oxford}
}

@article{10.1063/1.4858075,
    author = {Goryachev, Maxim and Farr, Warrick G. and Tobar, Michael E.},
    title = {Giant g-factors of natural impurities in synthetic quartz},
    journal = {Applied Physics Letters},
    volume = {103},
    number = {26},
    pages = {262404},
    year = {2013},
    month = {12},
    issn = {0003-6951},
    doi = {10.1063/1.4858075},
    url = {https://doi.org/10.1063/1.4858075}
}

@article{PhysRevApplied.21.064002,
  title = {Conductivity freeze-out in isotopically pure $\mathrm{Si}$-28 at millikelvin temperatures},
  author = {McAllister, Ben T. and Zhao, Zijun C. and Bourhill, Jeremy F. and Goryachev, Maxim and Creedon, Daniel and Johnson, Brett C. and Tobar, Michael E.},
  journal = {Phys. Rev. Appl.},
  volume = {21},
  issue = {6},
  pages = {064002},
  numpages = {11},
  year = {2024},
  month = {Jun},
  publisher = {American Physical Society},
  doi = {10.1103/PhysRevApplied.21.064002},
  url = {https://link.aps.org/doi/10.1103/PhysRevApplied.21.064002}
}

@article{10.1063/1.4920987,
    author = {Carvalho, N. C. and Le Floch, J-M. and Krupka, J. and Tobar, M. E.},
    title = {Multi-mode technique for the determination of the biaxial Y2SiO5 permittivity tensor from 300 to 6 K},
    journal = {Applied Physics Letters},
    volume = {106},
    number = {19},
    pages = {192904},
    year = {2015},
    month = {05},
    issn = {0003-6951},
    doi = {10.1063/1.4920987},
    url = {https://doi.org/10.1063/1.4920987}
}

@article{article1,
author = {C. Carvalho, Natália and Goryachev, Maxim and Krupka, Jerzy and Bushev, P. and Tobar, Michael},
year = {2017},
month = {02},
pages = {},
title = {Low Temperature Microwave Properties of Biaxial YAlO3},
volume = {96},
journal = {Physical Review B},
doi = {10.1103/PhysRevB.96.045141}
}

\end{document}